\documentclass[aps,prb,twocolumn,showpacs,preprintnumbers,amsmath,amssymb,superscriptaddress]{revtex4}

\usepackage{graphicx}
\usepackage{dcolumn}
\usepackage{bm}

\newcommand{\ket}[1]{|#1\rangle}

\newcommand{\matel}[3]{\langle #1|#2|#3 \rangle}
\newcommand{\e}{\varepsilon}

\begin{document}

\title{Effects of different geometries on the conductance,
shot noise and tunnel magnetoresistance of double quantum dots}

\author{Ireneusz Weymann}
\email{weymann@amu.edu.pl} \affiliation{Department of Physics,
Adam Mickiewicz University, 61-614 Pozna\'n, Poland}
\affiliation{Department of Theoretical Physics, Institute of
Physics, Budapest University of Technology and Economics, H-1521
Budapest, Hungary}

\date{\today}

\begin{abstract}
The spin-polarized transport through a coherent strongly coupled
double quantum dot (DQD) system is analyzed theoretically in the
sequential and cotunneling regimes. Using the real-time
diagrammatic technique, we analyze the current, differential
conductance, shot noise and tunnel magnetoresistance (TMR) as a
function of both the bias and gate voltages for double quantum
dots coupled in series, in parallel as well as for T-shaped
systems. For DQDs coupled in series, we find a strong dependence
of the TMR on the number of electrons occupying the double dot,
and super-Poissonian shot noise in the Coulomb blockade regime. In
addition, for asymmetric DQDs, we analyze transport in the Pauli
spin blockade regime and explain the existence of the leakage
current in terms of cotunneling and spin-flip cotunneling-assisted
sequential tunneling. For DQDs coupled in parallel, we show that
the transport characteristics in the weak coupling regime are
qualitatively similar to those of DQDs coupled in series. On the
other hand, in the case of T-shaped quantum dots we predict a
large super-Poissonian shot noise and TMR enhanced above the
Julliere value due to increased occupation of the decoupled
quantum dot. We also discuss the possibility of determining the
geometry of the double dot from transport characteristics.
Furthermore, where possible, we compare our results with existing
experimental data on nonmagnetic systems and find qualitative
agreement.
\end{abstract}

\pacs{72.25.Mk, 73.63.Kv, 85.75.-d, 73.23.Hk}

\maketitle

\section{Introduction}

Transport properties of double quantum dots (DQDs) have recently
attracted much interest.
\cite{gossardPRL95,hawrylakPRB95,blickPRB96,imamuraPRB98,zieglerPRB00,ono02,
derWielRMP03,hazalzetPRB01,golovachPRB04,graberPRB06,
cotaPRL05,mcclurePRL07,liuPRB05,wunschPRB06,
franssonPRB06,franssonNT06,aghassiPRB06,inarreaPRB07,
bartholdPRL06,kieslichPRL07,dattaPRB07,
pedersenPRB07,weymannPRB07,trochaPRB07} Since the behavior of
double quantum dots resembles the behavior of molecules, DQDs are
frequently called {\it artificial molecules}, and are thus
considered as ideal systems to study the fundamental many-body
interactions between single electrons and spins.
\cite{wolf01,loss02,maekawa02,zutic04,maekawa06} Double quantum
dots exhibit a variety of interesting effects, such as for example
current rectification due to the Pauli spin blockade,
\cite{imamuraPRB98,ono02,franssonPRB06,franssonNT06,inarreaPRB07,dattaPRB07}
negative differential conductance, \cite{liuPRB05} formation of
molecular states, \cite{hawrylakPRB95,blickPRB96,graberPRB06} spin
pumping, \cite{hazalzetPRB01,cotaPRL05} Kondo effect,
\cite{aguadoPRL00,lopezPRL02,chenPRL04} etc. Furthermore, double
quantum dots are also interesting for future applications in
quantum computing. \cite{lossPRA98,lossPRL00,huPRA00,hansonPRL07}
In addition, when the leads are ferromagnetic, transport
properties of the system strongly depend on the relative
orientation of the magnetizations of electrodes, leading to the
tunnel magnetoresistance (TMR) effect, spin accumulation, exchange
field, etc. \cite{julliere75,barnas98,bulka00,rudzinski01,
koenigPRL03,braunPRB04,weymannPRB05,weymannPRB07,trochaPRB07}

The problem of spin-polarized transport properties has been so far
mainly addressed in the case of single quantum dots.
\cite{bulka00,rudzinski01,koenigPRL03,braunPRB04,weymannPRB05,
weymannPRBBR05,weymannQDs,braigPRB05,souzaPRB07,franssonEPL05,
cottet04,cottetPRB06,kondoSQD,utsumiPRB05,sindelPRB07,weymannPRB08}
In particular, in the strong coupling regime it was shown that the
Kondo peak becomes split when the magnetic configuration changes
from the antiparallel to parallel one, and that this splitting can
be compensated upon applying external magnetic field.
\cite{kondoSQD,utsumiPRB05,sindelPRB07} In the weak coupling
regime, on the other hand, the parity effect of the linear
response TMR was predicted \cite{weymannPRB05} and the zero-bias
anomaly was found in the differential conductance when magnetic
moments of the leads form an antiparallel configuration.
\cite{weymannPRBBR05} In this paper we extend the existing
theoretical studies and consider transport through coherent double
quantum dots weakly coupled to external ferromagnetic leads. We
note that spin-dependent transport properties of DQDs have been
analyzed very recently in a few theoretical papers.
\cite{franssonNT06,weymannPRB07,hornberger07} The considerations
were however limited to either first-order tunneling or the
Coulomb blockade regime. In particular, in
Ref.~[\onlinecite{franssonNT06}] the formation of a pure triplet
state was predicted, Ref.~[\onlinecite{weymannPRB07}] deals with
transport in the deep Coulomb blockade regime, while
Ref.~[\onlinecite{hornberger07}] presents a detailed analysis of
sequential transport in the case of non-collinearly polarized
leads.

The goal of this paper is thus to analyze the transport properties
of double quantum dots coupled to collinear ferromagnetic leads in
the full weak coupling regime, i.e. including sequential,
cotunneling and cotunneling-assisted sequential processes.
Furthermore, we also analyze the effect of different geometries of
the double dots on transport, in particular, the cases of DQDs
coupled in series, in parallel, as well as T-shaped double quantum
dots are considered. The comparison of numerical results obtained
for different geometries may in principle help in determining the
system's geometry, which may be of importance in discussing and
understanding experimental results, especially on self-assembled
quantum dots. The present analysis is based on the real-time
diagrammatic technique which consists in a perturbation expansion
of the density matrix of the system under consideration, and the
relevant operators, with respect to the coupling to external
leads. The advantage of using the real-time diagrammatic technique
is that it takes into account the effects of the exchange field
and the renormalization of the dot levels in a fully systematic
way. In particular, in the case of asymmetric DQDs coupled in
series, we consider transport in the Pauli spin blockade regime
and show that the leakage current in the blockade region results
from cotunneling and cotunneling-assisted sequential tunneling
processes. As far as the shot noise is concerned, we show that the
noise is super-Poissonian in the Coulomb blockade regime and drops
to sub-Poissonian value in the sequential tunneling regime, where
the Fano factor approaches one half. This behavior is observed for
parallel and serial DQD's geometries. In turn, for T-shaped double
quantum dots, we find a large super-Poissonian shot noise due to
increased occupation of the decoupled dot. We also analyze the TMR
effect and find its strong dependence on the transport regime and
number of electrons in the DQD in the ground state. For DQDs
coupled in series and in parallel the TMR takes the values ranging
from around a half of the Julliere TMR to its full value,
\cite{julliere75} whereas for T-shaped DQDs the TMR may be
enhanced above the Julliere value.

The systems considered in this paper may be realized
experimentally for example in lateral and vertical semiconductor
quantum dots
\cite{ono02,pettaPRL04,mcclurePRL07,liuPRB05,johnsonPRB05} or
single wall metallic carbon nanotubes with top gate electrodes.
\cite{graberPRB06,jorgensenAPL06,sapmazNL06,graber06} The latter
systems are of particular interest because, by tuning the gates,
it is possible to change the charge on each dot separately.
Unfortunately, the aforementioned experiments concerned only DQDs
coupled to nonmagnetic leads. There are several experimental
realizations of single quantum dots attached to ferromagnetic
leads, \cite{chye02,heersche06,zhang05,tsukagoshi99,
zhao02,jensen05,sahoo05,pasupathy04,fertAPL06,hamayaAPL07a,hamayaAPL07b,
hamayaPRB08,parkinNL08} while experimental data on spin-polarized
transport through double quantum dots are lacking. We thus believe
that the results presented in this paper will be of assistance in
discussing future experiments.

The paper is organized as follows: The model of the considered
system is presented in Sec. II, whereas the method employed in
calculations is described in Sec. III. In Section IV we present
and discuss the numerical results for double quantum dots coupled
in series. In this section we also focus on the role of
second-order processes in transport. In addition, we also analyze
transport in the Pauli spin blockade regime. In Sec. V we deal
with DQDs coupled in parallel, whereas in Sec. VI we consider
transport through T-shaped quantum dots. Finally, the conclusions
are given in Sec. VII.

\section{Model}

\begin{figure}[t]
  \includegraphics[width=0.7\columnwidth]{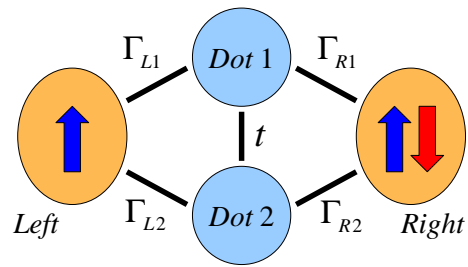}
  \caption{\label{Fig:1}
  (Color online) Schematic of a double quantum dot weakly coupled
  to external ferromagnetic electrodes.
  The magnetizations of the leads can
  form either parallel or antiparallel magnetic configuration.
  The first ($j=1$) and second ($j=2$) dots are coupled to each other via
  the hopping $t$, and to the left ($r=L$) and right ($r=R$) leads
  with the coupling strength $\Gamma_{rj}$. The double
  quantum dot is assumed to be symmetrically biased.
  By adjusting the couplings, the system can smoothly
  cross over from the serial to parallel geometry.}
\end{figure}

The schematic of a double quantum dot coupled to ferromagnetic
leads is shown in Fig.~\ref{Fig:1}. It is assumed that the
magnetizations of the leads are oriented collinearly, so that the
system can be either in the parallel or antiparallel magnetic
configuration. The Hamiltonian of the system is given by
\begin{equation}
  H=H_{\rm L} + H_{\rm R} + H_{\rm DQD} + H_{\rm T}
\end{equation}
The first two terms describe noninteracting itinerant electrons in
the leads, $H_r=\sum_{{\mathbf k}\sigma} \varepsilon_{r{\mathbf
k}\sigma} c^{\dagger}_{r{\mathbf k}\sigma} c_{r{\mathbf k}\sigma}$
for the left ($r={\rm L}$) and right ($r={\rm R}$) lead, where
$\varepsilon_{r{\mathbf k}\sigma}$ is the energy of an electron
with the wave vector ${\mathbf k}$ and spin $\sigma$ in the lead
$r$, and $c^{\dagger}_{r{\mathbf k}\sigma}$ ($c_{r{\mathbf
k}\sigma}$) denotes the respective creation (annihilation)
operator. The double dot is described by the Hamiltonian
\begin{eqnarray}\label{Eq:DQDHamiltonian}
  H_{\rm DQD} &=&
  \sum_{j=1,2}\sum_{\sigma}\varepsilon_{j} n_{j\sigma}
  + U \sum_{j=1,2} n_{j\uparrow}n_{j\downarrow}
  \nonumber\\
  &+& U^\prime \sum_{\sigma\sigma^\prime}
  n_{\rm 1\sigma}n_{\rm 2\sigma^\prime} +
  t \sum_\sigma(d^\dagger_{1\sigma}d_{2\sigma}+
  d^\dagger_{2\sigma}d_{1\sigma})
 \,,
\end{eqnarray}
with $n_{j\sigma}=d^{\dagger}_{j\sigma}d_{j\sigma}$, where
$d^{\dagger}_{j\sigma}$ ($d_{j\sigma}$) is the creation
(annihilation) operator of an electron with spin $\sigma$ in the
first ($j=1$) or second ($j=2$) quantum dot, and $\varepsilon_{j}$
is the corresponding single-particle energy. The Coulomb
interaction on the first and second dot is assumed to be equal and
is described by $U$, while $U^\prime$ corresponds to the inter-dot
Coulomb correlation. The last term of $\hat{H}_{\rm DQD}$
describes the hopping between the two dots with $t$ being the
hopping parameter. We assume that the hopping parameter is large,
so that there is a considerable overlap of the wave functions of
the two dots, leading to the formation of molecular many-body
states, through which transport takes place. In addition, we also
note that an exchange interaction between spins in the two dots
may lead to the formation of singlet and triplet states.
\cite{graberPRB06} However, experimentally, this exchange
interaction was found to be rather small as compared to the other
energy scales, \cite{ono02} therefore, in the following
considerations, we will neglect it.

The tunneling processes between the DQD and electrodes are
described by the Hamiltonian,
\begin{eqnarray}
  H_{\rm T} = \sum_{r={\rm L,R}}\sum_{j=1,2}\sum_{{\mathbf k}\sigma}
  \left(
    t_{rj} c^{\dagger}_{r {\mathbf k}\sigma} d_{j \sigma} +
    t^\star_{rj} d^\dagger_{j \sigma} c_{r {\mathbf k}\sigma}
  \right)
   \,,
\end{eqnarray}
where $t_{rj}$ denotes the tunnel matrix elements between the
$r$th lead and the $j$th dot. The coupling of the $j$th dot to the
$r$th lead can be written as $\Gamma_{rj}^{\sigma}= 2\pi
|t_{rj}|^2 \rho_r^\sigma$, where $\rho_r^\sigma$ is the
spin-dependent density of states of lead $r$. By introducing the
definition of the spin polarization of lead $r$,
$p_{r}=(\rho_{r}^{+}- \rho_{r}^{-})/ (\rho_{r}^{+}+
\rho_{r}^{-})$, one can write,
$\Gamma_{rj}^{+(-)}=\Gamma_{rj}(1\pm p_{r})$, with $\Gamma_{rj}=
(\Gamma_{rj}^{+} +\Gamma_{rj}^{-})/2$, where $\Gamma_{rj}^{+}$ and
$\Gamma_{rj}^{-}$ describe the coupling of the $j$th dot to the
spin-majority and spin-minority electron bands of the lead $r$,
respectively.

Because the double dot Hamiltonian, Eq.~(\ref{Eq:DQDHamiltonian}),
is not diagonal in the local basis, we perform a unitary
transformation, $U^\dagger H_{\rm DQD} U = \tilde{H}_{\rm DQD}$,
to a new basis in which $\tilde{H}_{\rm DQD}$ is diagonal,
$\tilde{H}_{\rm DQD} \ket{\chi} = \varepsilon_\chi \ket{\chi}$.
The eigenvectors $\ket{\chi}$ are the many-body states of the
double quantum dot, while the eigenvalues $\varepsilon_\chi$
denote the corresponding energies. \cite{kostyrkoPRB04}

\section{Method}

In order to calculate the spin-dependent transport properties of a
double quantum dot in the sequential and cotunneling regimes, we
employ the real-time diagrammatic technique.
\cite{weymannPRB05,diagrams,thielmann} It consists in a systematic
perturbation expansion of the reduced density matrix of the
considered system and the current operator with respect to the
dot-lead coupling strength $\Gamma$. The time evolution of the
reduced density matrix can be visualized as a sequence of
irreducible self-energy blocks, $W_{\chi \chi^\prime}$, on the
Keldysh contour, where $W_{\chi \chi^\prime}$ describe transitions
between the many-body DQD states $\ket{\chi}$ and
$\ket{\chi^\prime}$. The elements $W_{\chi \chi^\prime}$ set up a
self-energy matrix $\mathbf{W}$. Within the matrix notation
introduced by Thielmann {\it et al.}, \cite{thielmann} the
stationary elements of the reduced density matrix can be found
from
\begin{equation}
  (\mathbf{\tilde{W}}\mathbf{p}^{\rm st})_{\chi} =
  \Gamma\delta_{\chi\chi_0}\,,
\end{equation}
where $\mathbf{p}^{\rm st}$ is the vector containing probabilities
and the matrix $\mathbf{\tilde{W}}$ is the modified matrix
$\mathbf{W}$ so as to include the normalization of probabilities.
The current flowing through the system can be then found from
\begin{equation}\label{Eq:current}
  I=\frac{e}{2\hbar}{\rm Tr}\{\mathbf{W}^{\rm I}\mathbf{p}^{\rm st}\}
  \,.
\end{equation}
The matrix $\mathbf{W}^{\rm I}$ is the self-energy matrix with one
{\it internal} vertex resulting from the expansion of the
tunneling Hamiltonian replaced by the current operator.

To calculate the transport properties order by order in tunneling
processes, we expand the self-energy matrices, $\mathbf{W}=
\mathbf{W}^{(1)} +\mathbf{W}^{(2)}+\dots$, $\mathbf{W}^{\rm I} =
\mathbf{W}^{\rm I(1)} +\mathbf{W}^{\rm I(2)}+\dots$, and the dot
occupations, $\mathbf{p}^{\rm st} = \mathbf{p}^{\rm st (0)} +
\mathbf{p}^{\rm st (1)}+\dots$, respectively. Then, the
first-order (sequential) and the second-order (cotunneling)
currents are given by
\begin{eqnarray}\label{Eq:I1I2}
  I^{(1)}&=&\frac{e}{2\hbar}{\rm Tr}\{\mathbf{W}^{\rm I (1)}\mathbf{p}^{\rm st (0)}\}
    \nonumber\,, \\
  I^{(2)}&=&\frac{e}{2\hbar}{\rm Tr}\{\mathbf{W}^{\rm I (2)}\mathbf{p}^{\rm st (0)}+
    \mathbf{W}^{\rm I (1)}\mathbf{p}^{\rm st (1)}\}  \,.
\end{eqnarray}
On the other hand, the zeroth and first-order occupation
probabilities can be found from the following equations
\begin{eqnarray}
  (\mathbf{\tilde{W}}^{(1)}\mathbf{p}^{\rm st (0)})_{\chi} &=&
  \Gamma\delta_{\chi\chi_0}\nonumber \,,\\
  \mathbf{\tilde{W}}^{(1)}\mathbf{p}^{\rm st (1)}+
  \mathbf{\tilde{W}}^{(2)}\mathbf{p}^{\rm st (0)} &=& 0\,,
\end{eqnarray}
where $\mathbf{\tilde{W}}^{(1)}$
$\left[\mathbf{\tilde{W}}^{(2)}\right]$ is given by
$\mathbf{W}^{(1)}$ $\left[\mathbf{W}^{(2)}\right]$ with one
arbitrary row $\chi_0$ replaced by $(\Gamma,\dots,\Gamma)$
$[(0,\dots,0)]$ due to the normalization of probabilities,
$\sum_{\chi}p_{\chi}^{\rm st (n)}=\delta_{n,0}$.

As one can see from the above formulas, to determine the transport
properties it is necessary to calculate the elements of the
corresponding self-energy matrices. This can be done using the
respective diagrammatic rules.
\cite{thielmann,diagrams,weymannPRB05} Although the first-order
calculation is rather simple, the general analytical formulas for
the self-energies in the second order are rather complicated due
to many virtual states through which the cotunneling processes can
take place. In the Appendix A, as an example, we present the
contribution coming from $W_{\chi(N),\chi^\prime(N)}^{(2)}$ where
$N$ is the charge state of the double dot.

In addition, in the present paper we also analyze the
zero-frequency current noise of the double quantum dot system,
$S=2\int_{-\infty}^0 dt (\langle
\hat{I}(t)\hat{I}(0)+\hat{I}(0)\hat{I}(t)\rangle-2 \langle
\hat{I}\rangle^2)$. For low bias voltages the current noise is
dominated by thermal noise, while for $|eV|>k_{\rm B}T$ the noise
associated with the discrete nature of charge (shot noise)
dominates. \cite{blanterPR00} Within the real-time diagrammatic
technique, a general formula for the shot noise has been derived
in Ref.~[\onlinecite{thielmann}] taking into account the
non-Markovian effects. The shot noise in the first-order can be
found from the corresponding expression
\begin{equation}\label{Eq:noise1}
  S^{(1)} = \frac{e^2}{\hbar}{\rm Tr}\left[\left(
  \mathbf{W}^{\rm II(1)}+
  \mathbf{W}^{\rm I(1)}\mathbf{P}^{(-1)}\mathbf{W}^{\rm I(1)}\right)
  \mathbf{p}^{\rm st(0)}
  \right] \,,
\end{equation}
while the cotunneling current noise is given by
\begin{eqnarray}\label{Eq:noise2}
  S^{(2)} &=& \frac{e^2}{\hbar}{\rm Tr}\left\{\left[
  \mathbf{W}^{\rm II(2)}+
  \mathbf{W}^{\rm I(2)}\mathbf{P}^{(-1)}\mathbf{W}^{\rm I(1)}
  \right.\right.\nonumber\\
  &+&\mathbf{W}^{\rm I(1)}\mathbf{P}^{(-1)}\mathbf{W}^{\rm I(2)}
  +\mathbf{W}^{\rm I(1)}\mathbf{P}^{(0)}\mathbf{W}^{\rm
  I(1)} \nonumber\\
  &+&\left.
  \mathbf{W}^{\rm I(1)}\mathbf{Q}^{(0)}
  \partial\mathbf{W}^{\rm I(1)}\right]
  \mathbf{p}^{\rm st(0)}\nonumber\\
  &+&\left. \left[ \mathbf{W}^{\rm II(1)}+
  \mathbf{W}^{\rm I(1)}\mathbf{P}^{(-1)}\mathbf{W}^{\rm I(1)}
  \right]  \mathbf{p}^{\rm st(1)} \right\} \,,
\end{eqnarray}
where $\mathbf{Q}^{(n)} = \mathbf{p}^{{\rm st}(n)}\otimes
\mathbf{e}^{\rm T}$, with $\mathbf{e}^{\rm T} = (1,\dots,1)$. The
objects $\mathbf{P}^{(-1)}$ and $\mathbf{P}^{(0)}$ are given by,
$\mathbf{\tilde{W}}^{(1)}\mathbf{P}^{(-1)} = \mathbf{Q}^{(1)} -
\mathbf{1}$, and, $\mathbf{\tilde{W}}^{(1)}\mathbf{P}^{(0)} +
\mathbf{\tilde{W}}^{(2)}\mathbf{P}^{(-1)} =
\mathbf{\tilde{1}}(\mathbf{Q}^{(1)}-\partial\mathbf{W}^{(1)}
\mathbf{Q}^{(0)})$, respectively, with $\mathbf{\tilde{1}}$ being
the unit vector with row $\chi_0$ set to zero. On the other hand,
the matrices $\mathbf{W}^{\rm II(1)}$ and $\mathbf{W}^{\rm II(2)}$
are the first and second-order self-energy matrices with two
internal vertices replaced by the current operator, while
$\partial\mathbf{W}^{(1)}$ and $\partial\mathbf{W}^{\rm I(1)}$ are
partial derivatives of $\mathbf{W}^{(1)}$ and $\mathbf{W}^{\rm
I(1)}$ with respect to the convergence factor of the Laplace
transform. \cite{thielmann}

By taking into account all the first-order and second-order
contributions to the self-energies, we are able to resolve the
transport properties in the full range of the bias and gate
voltages. The first order of expansion with respect to the
coupling corresponds to sequential tunneling, while the second
order is associated with cotunneling. Sequential tunneling is
allowed if the applied bias is larger than the threshold voltage,
i.e. when the energy provided by the transport voltage is
comparable with the charging energy. Otherwise, the system is in
the Coulomb blockade regime where sequential tunneling is
exponentially suppressed and the current flows due to cotunneling,
which involves correlated tunneling through virtual states of the
system. \cite{nazarov,kang,averin92}

Among different cotunneling processes one can generally
distinguish two types of processes: non-spin-flip ones, which do
not affect the DQD state, and the spin-flip ones. Furthermore, one
can also have double-barrier cotunneling, which contributes
directly to the current, and single-barrier cotunneling, which
affects the double dot occupations and, thus, indirectly, the
current.

\section{Double quantum dots coupled in series}

In this section we present and discuss numerical results on double
quantum dots coupled in series. In order to realize this geometry
we set $\Gamma_{{\rm L}2} = \Gamma_{{\rm R}1} = 0$ and assume
$\Gamma_{{\rm L}1} = \Gamma_{{\rm R}2} \equiv \Gamma/2$, see
Fig.~\ref{Fig:1}. Furthermore, in the following we will also
distinguish between symmetric and asymmetric DQDs. In the former
case, the level position of each dot is the same, $\varepsilon_1 =
\varepsilon_2$, while in the latter case the levels are detuned,
$\varepsilon_1 \neq \varepsilon_2$. In addition, in this section
we will emphasize the role of second-order processes in transport.
In order to ascribe observed features to respective tunneling
processes, we will therefore also present results obtained within
the sequential tunneling approximation, i.e. when considering only
the first-order tunneling processes.

\subsection{Symmetric double quantum dots}

\begin{figure}[t]
  \includegraphics[width = 0.7\columnwidth]{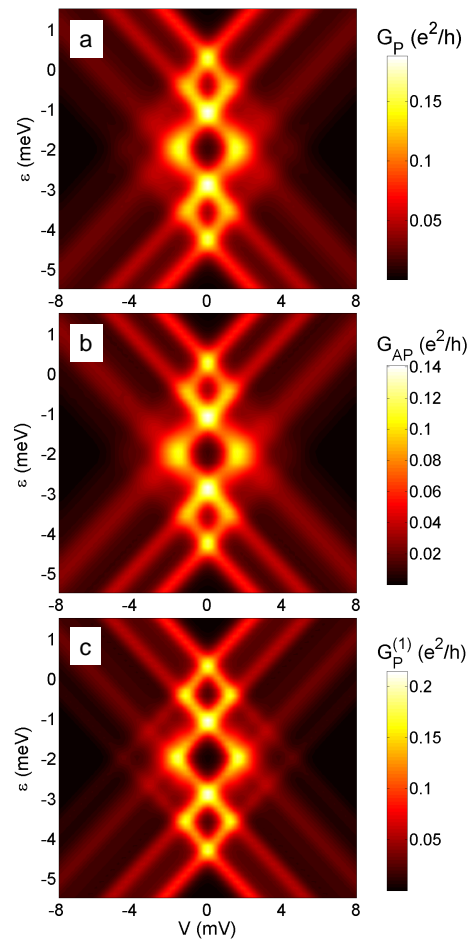}
  \caption{\label{Fig:2} (Color online)
  The differential conductance in the parallel $G_{\rm P}$
  (a) and antiparallel $G_{\rm AP}$ (b) magnetic configurations
  as a function of the bias voltage $V$
  and the position of the dots' levels
  $\varepsilon \equiv \varepsilon_1 = \varepsilon_2$
  for double quantum dots coupled in series.
  The parameters are:
  $k_{\rm B} T=0.15$ meV, $U=2$ meV, $U^\prime=1$ meV,
  $t=0.25$ meV,
  $\Gamma_{{\rm L}2} = \Gamma_{{\rm R}1} = 0$,
  $\Gamma_{{\rm L}1} = \Gamma_{{\rm R}2} \equiv \Gamma/2$, with
  $\Gamma=0.1$ meV, and $p=0.5$. Because the double quantum dot
  is symmetric, $\varepsilon_1 = \varepsilon_2$,
  there is no asymmetry associated with the bias reversal.
  For comparison in part (c) we also show the density plot
  of the differential conductance in the parallel configuration
  calculated using only the first-order tunneling processes, $G_{\rm P}^{(1)}$.}
\end{figure}

The differential conductance as a function of the bias voltage and
the position of the dot levels for the parallel and antiparallel
configurations in shown in Fig.~\ref{Fig:2}(a) and (b). Because
experimentally the position of the dots' levels can be changed
upon applying a gate voltage, Fig.~\ref{Fig:2} effectively shows
the bias and gate voltage dependence of the conductance. First of
all, it can be seen that the differential conductance displays a
characteristic Coulomb diamond structure. The diamonds in the low
bias voltage regime correspond to the Coulomb blockade regime
where the sequential current is exponentially suppressed and the
current flows due to cotunneling. With increasing the bias voltage
the excited states start participating in transport, which leads
to additional lines in the differential conductance. The features
described above are rather associated with the energy spectrum and
charge states of the DQD than with the ferromagnetism of the
leads, therefore they are present in both magnetic configurations.
The spin dependence stemming from ferromagnetic electrodes gives
rise to a difference of the conductance in the parallel and
antiparallel configurations, see Fig.~\ref{Fig:2}(a) and (b). The
conductance in the parallel configuration is generally larger than
that in the antiparallel configuration. This is due to the spin
asymmetry in the couplings when the leads' magnetizations are
antiparallel. As in the parallel configuration the majority
(minority) electrons of, let us say, left lead tunnel to the
majority (minority) electron band of the right lead, in the
antiparallel configuration the situation is reversed -- the
majority (minority) electrons of the left lead tunnel to the
minority (majority) electron band of the right lead. This leads to
the difference of the conductance in both magnetic configurations,
see Fig.~\ref{Fig:2}, and to the corresponding tunnel
magnetoresistance effect. In addition, in Fig.~\ref{Fig:2}(c) we
also present the bias and gate voltage dependence of the
differential conductance for the parallel configuration calculated
within the sequential tunneling approximation, $G_{\rm P}^{(1)}$.
By comparing Fig.~\ref{Fig:2}(a) and (c), one can see that
cotunneling leads to broadening of the resonance peaks, lowering
the magnitude of the differential conductance, and gives rise to
finite conductance in the Coulomb blockade regimes.

\begin{figure}[t]
  \includegraphics[width = 0.7\columnwidth]{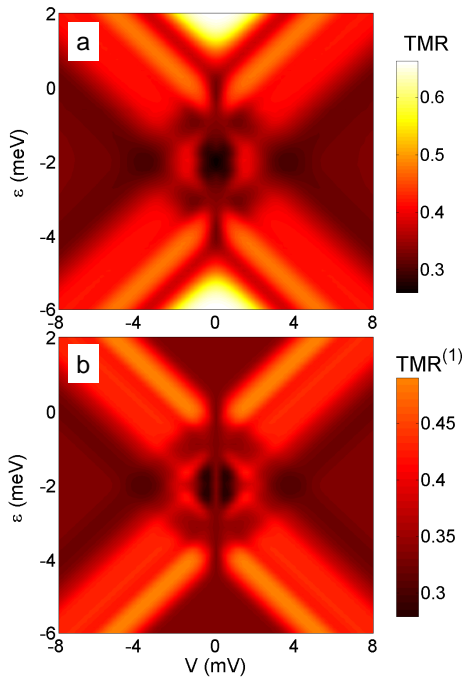}
  \caption{\label{Fig:3} (Color online)
  The total (first and second order) TMR (a)
  and the first-order TMR (b) as a function
  of the bias voltage $V$ and the level position $\varepsilon$
  for parameters the same as in Fig.~\ref{Fig:2}.
  The two figures are plotted in the same scale.}
\end{figure}

\begin{figure}[t]
  \includegraphics[width = 0.7\columnwidth]{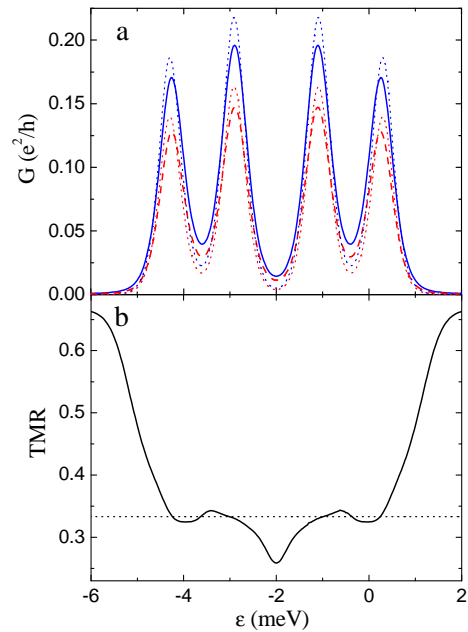}
  \caption{\label{Fig:4} (Color online)
  The linear conductance in the parallel (solid line)
  and antiparallel (dashed line) magnetic configurations (a)
  and the linear TMR (b) as a function
  of the level position $\varepsilon$
  for parameters the same as in Fig.~\ref{Fig:2}.
  For comparison, the dotted curves
  show the linear conductance and the TMR calculated
  taking into account only the first-order processes.}
\end{figure}

\begin{figure}[t]
  \includegraphics[height=8.7cm]{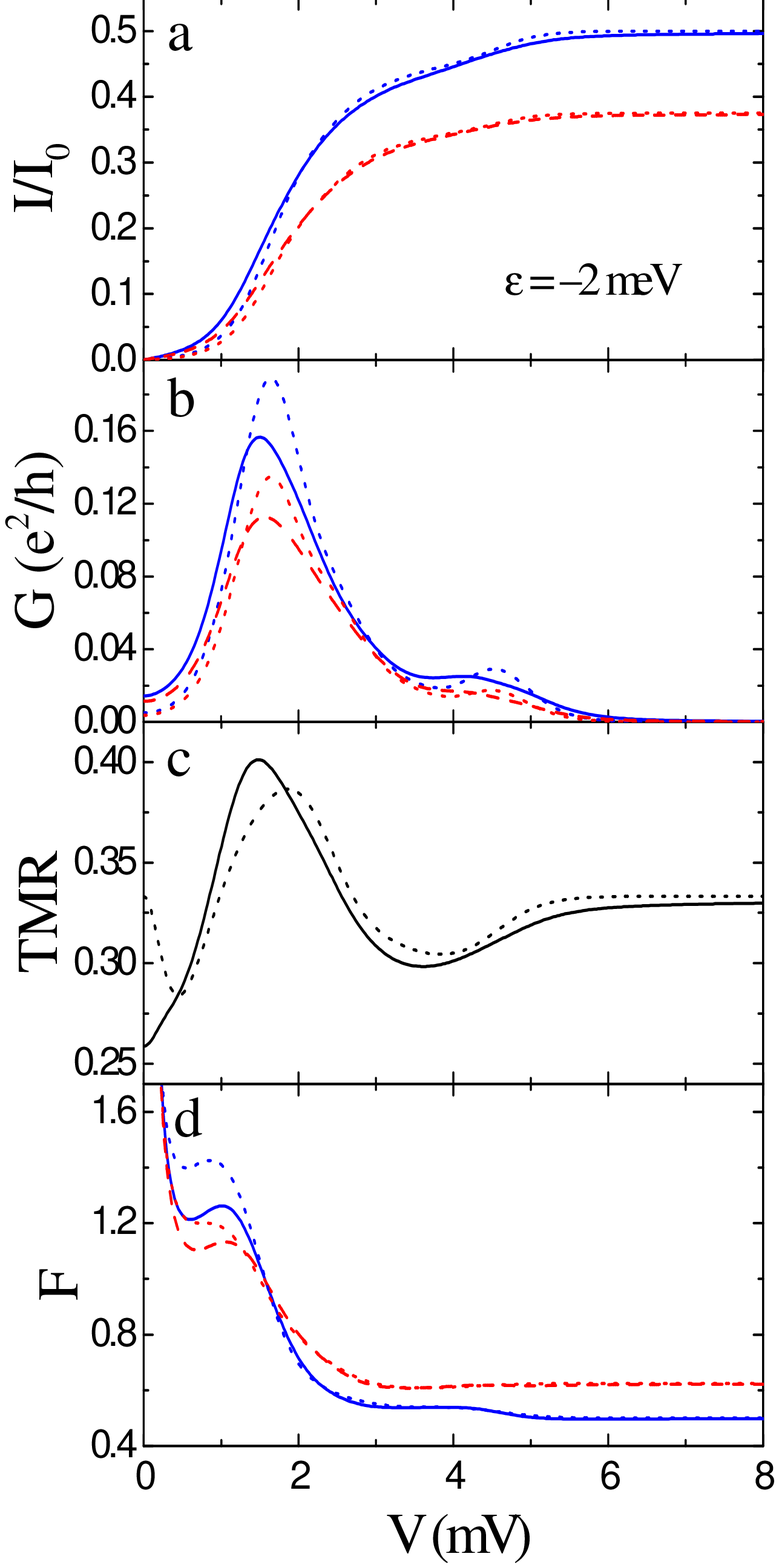}
  \includegraphics[height=8.7cm]{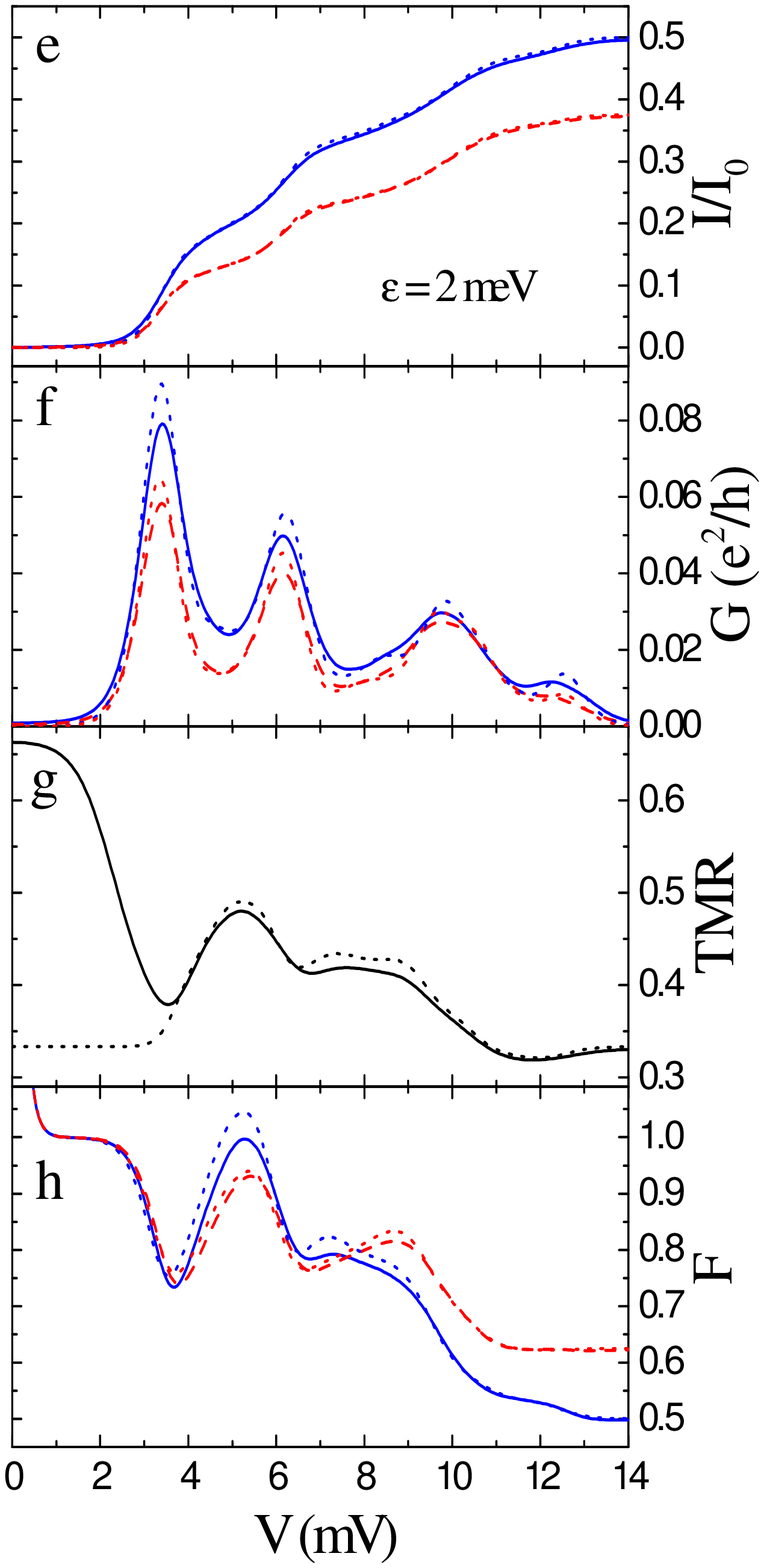}
  \caption{\label{Fig:5} (Color online)
  The current in units of $I_0 = e\Gamma/\hbar$ (a,e),
  differential conductance (b,f)
  and the Fano factor (d,h) in
  the parallel (solid line) and antiparallel configuration
  (dashed line) as well as the TMR (c,g)
  as a function of the bias voltage for
  different values of the level position $\varepsilon$
  as indicated in the figure. $\varepsilon=-2$ meV (a-d)
  corresponds to the case when the ground state of the DQD
  is doubly occupied, while for $\varepsilon=2$ meV (e-h)
  the double dot is empty.
  The parameters are the same as in Fig.~\ref{Fig:2}.
  The dotted curves show the results obtained in the
  sequential tunneling approximation.}
\end{figure}

The bias and gate voltage dependence of the TMR is shown in
Fig.~\ref{Fig:3}(a), while Fig.~\ref{Fig:3}(b) presents the
first-order TMR, ${\rm TMR}^{(1)}$. Intuitively, the TMR is a
measure of the system's transport properties change when the
magnetic configuration switches from the parallel to the
antiparallel one. It is defined as
\cite{barnas98,julliere75,weymannPRB05}
\begin{equation}
  {\rm TMR} = \frac{I_{\rm P}-I_{\rm AP}}{I_{\rm AP}} \,,
\end{equation}
where $I_{\rm P}$ ($I_{\rm AP}$) is the current flowing through
the system in the parallel (antiparallel) configuration. By
comparison with Fig.~\ref{Fig:2}, one can easily identify the
different transport regimes. Furthermore, by comparing the total
and sequential TMR, one can immediately see that cotunneling
modifies the TMR mainly in the low bias voltage regime, i.e. in
the Coulomb blockade regime and in the regime where sequential
tunneling is suppressed due to the absence of levels in the energy
window provided by transport voltage. In order to discuss and see
more clearly the behavior of the TMR on applied voltages, in
Fig.~\ref{Fig:4} we show the linear response conductance and
resulting TMR, while in Fig.~\ref{Fig:5} we display the transport
properties in the nonlinear response regime.

The linear conductance in both magnetic configurations is shown in
Fig.~\ref{Fig:4}(a). When lowering the position of the DQD levels
the conductance displays four resonance peaks associated with
subsequent occupation of the corresponding charge states. The
dotted curves in Fig.~\ref{Fig:4} present results obtained in the
first-order approximation. It can be seen that off resonance the
current is mainly dominated by cotunneling. On the other hand, on
resonance the sequential processes dominate current, while
cotunneling only slightly affects the conductance, leading to
renormalization of the DQD levels and, thus, slightly shifting the
position of the conductance peaks, see Fig.~\ref{Fig:4}(a).
Interestingly, the second-order processes have a rather large
impact on the linear TMR shown in Fig.~\ref{Fig:4}(b). First of
all, the linear response TMR exhibits a strong dependence on the
gate voltage, i.e. on the number of electrons in the double
quantum dot. When the DQD is either empty or fully occupied, the
TMR is given by the Julliere value, ${\rm TMR} = 2p^2 / (1-p^2)$.
\cite{julliere75} This is due to the fact that in those transport
regimes the current is driven by elastic cotunneling processes
which do not affect the DQD state in any way. Thus, as far as the
TMR is concerned, the system behaves as a single ferromagnetic
tunnel junction. \cite{julliere75} However, in the Coulomb
blockade regions, the TMR becomes generally suppressed due to the
presence of inelastic cotunneling which introduces spin-flip
processes in the system. The dotted line in Fig.~\ref{Fig:4}(b)
shows the TMR calculated within the sequential tunneling
approximation -- it is given by ${\rm TMR} = p^2/(1-p^2)$.
\cite{weymannPRB05,hornberger07} At this point, we would like to
note that the results for the linear conductance calculated by
including only the first-order processes are rather reliable in
the whole range of $\varepsilon$. On the other hand, the results
for the linear response TMR are comparable to those obtained
within the sequential tunneling approximation only on resonance,
where sequential processes dominate, while off resonance they are
completely unreliable, see Fig. ~\ref{Fig:4}. Therefore, in order
to properly analyze the dependence of the TMR in the full range of
parameters one has to take into account cotunneling processes.

The bias voltage dependence of the current, differential
conductance, tunnel magnetoresistance and the Fano factor is shown
in Fig.~\ref{Fig:5} for $\varepsilon=-2$ meV and $\varepsilon=2$
meV. The first situation, $\varepsilon=-2$ meV, corresponds to the
case when the ground state of the double quantum dot is doubly
occupied, while the second one, $\varepsilon=2$ meV, corresponds
to the case of empty double dot. In all cases, due to the spin
asymmetry of tunneling processes, the current in the parallel
configuration is larger than the current in the antiparallel
configuration. In addition, the i-v curves display characteristic
Coulomb steps. With increasing the bias voltage more and more
charge states start to participate in transport which gives rise
to the corresponding steps in the current-voltage characteristics,
and peaks in the differential conductance. Furthermore, the
single-electron charging effects also lead to the oscillatory-like
behavior of the TMR effect, see Fig.~\ref{Fig:5}(c) and (g). In
the Coulomb blockade regime, $\varepsilon=-2$ meV, the TMR is much
suppressed as compared to the case of $\varepsilon=2$ meV. This is
due to the presence of spin-flip cotunneling processes in the case
of the doubly occupied DQD, as already discussed in association
with Fig.~\ref{Fig:4}. However, when increasing the bias voltage
the TMR increases and, at the threshold for sequential tunneling,
reaches a local maximum, see Fig.~\ref{Fig:5}(c). This effect is
due to the nonequilibrium spin accumulation which is induced in
the DQD system with increasing the transport voltage. (Similar
effect can be observed when the double dot is singly occupied,
i.e. when $\varepsilon = -0.5$ meV.) On the other hand, when the
DQD is empty, the TMR at low bias voltage acquires the Julliere
value, \cite{julliere75} and then, with increasing $V$, becomes
generally suppressed, see Fig.~\ref{Fig:5}(g).

\begin{figure}[t]
  \includegraphics[width = 0.7\columnwidth]{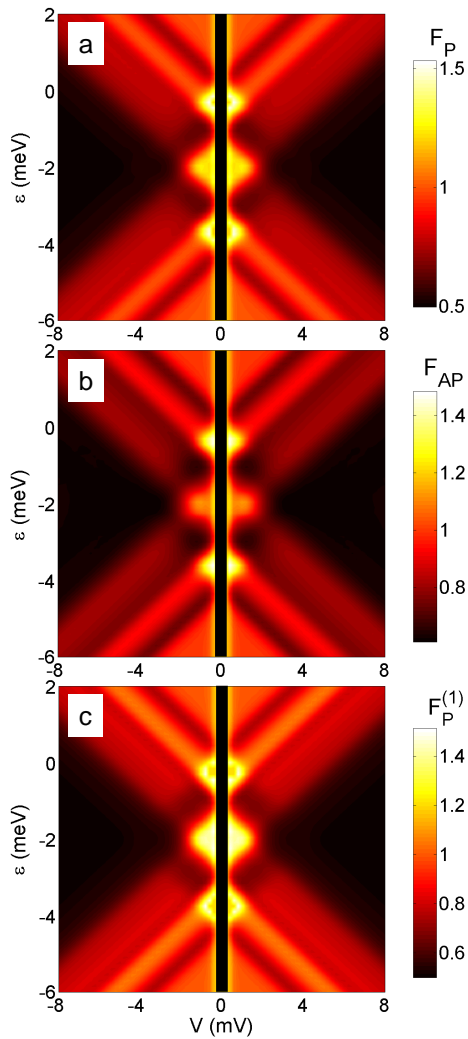}
  \caption{\label{Fig:6} (Color online)
  The total Fano factor in the parallel (a)
  and antiparallel (b) magnetic configurations
  as a function of the bias voltage $V$
  and the position of the dots' levels
  $\varepsilon$ for parameters the same as in Fig.~\ref{Fig:2}.
  Part (c) presents the Fano factor in the
  parallel configuration calculated in
  the first-order approximation.
  Because the Fano factor is divergent
  for $|eV|\lesssim k_{\rm B}T$, this transport regime
  is denoted with a thick black line.}
\end{figure}

In addition, in Fig.~\ref{Fig:5} we also show the Fano factor,
$F=S/(2e|I|)$, calculated in both magnetic configurations and for
two values of $\varepsilon$, as indicated in the figure. First of
all, one can see that the Fano factor becomes divergent at low
bias voltages. This is due to the thermal noise which dominates
the current noise as $V\to 0$, while $I\to 0$, leading to
$F\to\infty$. \cite{blanterPR00} Furthermore, it can be seen that
the Fano factor in the Coulomb blockade regime is slightly larger
in the parallel configuration than in the antiparallel one. This
is associated with an additional positive cross-correlations,
which contribute to the current noise, due to the ferromagnetism
of the electrodes. In the Coulomb blockade regime, there is a
larger asymmetry between the cotunneling processes through the
majority and minority spin channels in the parallel configuration
than in the antiparallel one. This effectively increases the
current fluctuations in the parallel configuration and leads to
the corresponding difference in the Fano factors, see
Fig.~\ref{Fig:5}(d). On the other hand, in the sequential
tunneling regime, the information about the spin asymmetry of
tunneling processes between the DQD and the corresponding lead is
rather contained in the nonequilibrium spin accumulation induced
in the double dot. This spin accumulation is larger in the
antiparallel configuration, yielding generally an enhanced Fano
factor in the antiparallel configuration as compared to the
parallel one. Moreover, we also note that in the Coulomb blockade
regime, $\varepsilon=-2$ meV, the shot noise becomes
super-Poissonian ($F>1$) and drops to the sub-Poissonian value
($F<1$) at the threshold for sequential tunneling, see
Fig.~\ref{Fig:5}(d). This super-Poissonian shot noise is
associated with bunching of inelastic cotunneling processes
through the system. \cite{weymannPRB08,sukhorukovPRB01} On the
other hand, in the case of empty DQD, $\varepsilon=2$ meV, in the
cotunneling regime the current flows due to elastic second-order
processes which obey the Poissonian statistics. In this case the
shot noise is given by $S=2e|I|$ and the Fano factor is simply
equal to unity, see Fig.~\ref{Fig:5}(h), irrespective of magnetic
configuration of the system. As concerns the sequential tunneling
regime, the shot noise is then generally sub-Poissonian,
irrespective of the position of the DQD level $\varepsilon$. This
indicates the role of correlations in electronic transport, in
particular, the Coulomb correlations and charge conservation.
\cite{blanterPR00} Our results are in qualitative agreement with
experimental data on tunneling through vertically coupled
self-assembled InAs quantum dots where also super-Poissonian shot
noise has been found. \cite{bartholdPRL06}

Finally, to make the present analysis self-contained, we also show
the bias and gate voltage dependence of the total Fano factor for
the parallel and antiparallel magnetic configuration, see
Fig.~\ref{Fig:6}(a) and (b), respectively. In addition, for
comparison, the Fano factor calculated by taking into account only
first-order tunneling processes is depicted in
Fig.~\ref{Fig:6}(c). The black lines around the zero bias,
$|eV|\lesssim k_{\rm B}T$, mark the transport regime where the
Fano factor is divergent due to finite thermal noise. The
different behavior of the shot noise is now clearly visible. In
the Coulomb blockade regime the noise is super-Poissonian, in the
cotunneling regime when the DQD is either empty of fully occupied
(current is mediated by elastic cotunneling) the shot noise is
Poissonian, and in the sequential tunneling regime the noise drops
to sub-Poissonian value.

\subsection{Asymmetric double quantum dots: Pauli spin blockade}

By applying a gate voltage to each quantum dot, it is possible to
tune the dot levels separately. So far, we have considered the
case of symmetric DQD, i.e. when $\varepsilon_1 = \varepsilon_2$.
In this situation the transport characteristics were symmetric
with respect to the bias reversal. This is however not the case
for $\varepsilon_1 \neq \varepsilon_2$, where the current becomes
asymmetric with respect to the bias reversal, leading to the Pauli
spin blockade and negative differential conductance, as observed
experimentally. \cite{ono02,liuPRB05} In the following, assuming
realistic parameters of the double quantum dot system, we analyze
transport properties in the regime where the Pauli spin blockade
effects are visible.

\begin{figure}[t]
  \includegraphics[width = 0.68\columnwidth]{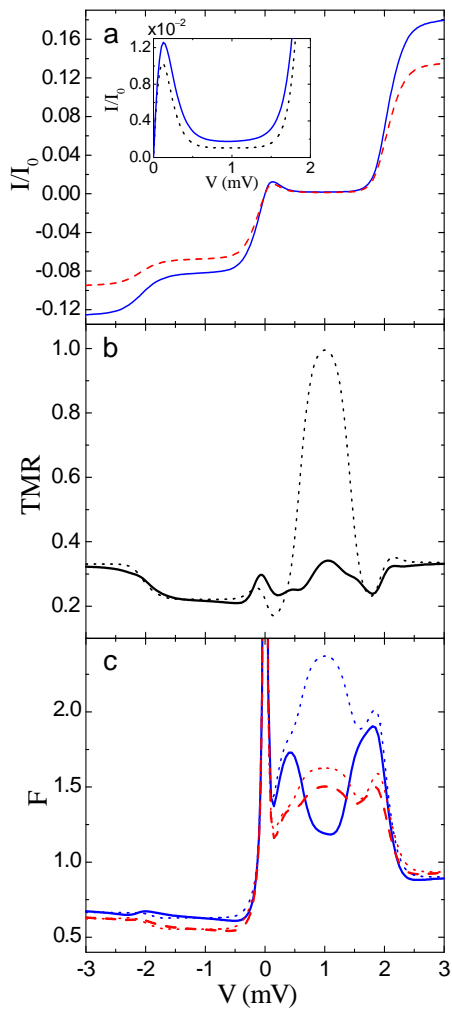}
  \caption{\label{Fig:7} (Color online)
  The current (a) and Fano factor (c)
  in the parallel (solid) and antiparallel (dashed)
  magnetic configurations as well
  as the TMR (b) as a function of the bias voltage for DQD coupled in series.
  The parameters are:
  $k_{\rm B} T=0.05$ meV,
  $\varepsilon_1 = -1$ meV, $\varepsilon_2 = -2$ meV,
  $\Delta = 3$ meV,
  $U=2$ meV, $U^\prime=1$ meV,
  $t=0.05$ meV, where $\Delta$ is the level spacing in each dot.
  The other parameters are the same as in Fig.~\ref{Fig:2}.
  The inset in part (a) displays the current in the
  parallel magnetic configuration in the Pauli spin blockade regime.
  The dotted lines correspond to the first-order calculation.}
\end{figure}

The mechanism leading to the spin blockade was theoretically
discussed by Fransson {\it et al.} \cite{franssonPRB06} and
Muralidharan {\it et al.} \cite{dattaPRB07} These considerations
were however restricted only to the first-order tunneling
processes, which dominate the current out of the blockade regime.
As shown experimentally by Ono {\it et al.}, \cite{ono02} in the
spin blockade regime there is a finite leakage current, which
cannot be explained within the sequential tunneling approximation.
To explain the existence of this leakage, it was proposed that
non-zero current in the Pauli blockade is associated with
spin-flip processes induced by hyperfine interaction.
\cite{inarreaPRB07} In the following, we show that the leakage
current results just from the interplay between different
intrinsic tunneling processes driving the current. The hyperfine
interaction, or coupling to a phonon bath, may, of course,
increase the leakage, but is not necessary for the observation of
a finite current in the Pauli spin blockade regime.

When the DQDs levels are detuned, $\varepsilon_2 < \varepsilon_1$,
and once the first and second dot become singly occupied, the
current may be suppressed in some range of the bias voltage. This
is associated with the full occupation of two-electron triplet
states of the DQD. \cite{franssonPRB06} The Pauli spin blockade
may be lifted when the applied bias voltage admits another
(exited) charge states to participate in transport. On the other
hand, when the voltage is reversed, it is energetically allowed
that the electron from the first dot tunnels to the left lead and
then another electron from the right lead enters the DQD, lifting
the Pauli spin blockade.

This can be seen in Fig.~\ref{Fig:7} where we present the bias
voltage dependence of the current, Fano factor and the TMR effect.
In order to make the calculations more realistic and to allow for
excited states of the system, we have now taken into account four
different orbital levels, two in each dot. Such system is
described by straightforward extension of the hamiltonian,
Eq.~(\ref{Eq:DQDHamiltonian}), where the level spacing in the dots
is described by the parameter $\Delta$. First of all, one can see
that the transport characteristics are asymmetric with respect to
the bias reversal. This is associated with the fact that by
detuning the DQDs levels the symmetry of transition rates between
the left and right leads has been broken. Furthermore, for
positive bias voltage the current is suppressed in a broad range
of the bias voltage ($0<eV<U$) due to the full occupation of the
triplet states, while for negative bias the Pauli spin blockade is
lifted. In addition, the blockade can also be lifted when the
positive bias voltage is increased further, $eV>U$, so that
tunneling through the second level of the second dot is allowed,
see Fig.~\ref{Fig:7}(a).

In the inset of Fig.~\ref{Fig:7}(a) we show the current in the
parallel configuration just in the spin blockade regime. For
comparison, we also display the current calculated within the
sequential tunneling approximation. One can see that the
sequential current is suppressed as compared to the total current
(calculated taking into account cotunneling processes). The
suppression of the sequential current is generally governed by the
ratio $2t/|\varepsilon_2-\varepsilon_1|$. \cite{franssonPRB06}
When $2t/|\varepsilon_2-\varepsilon_1| \ll 1$, the occupation of
the two-electron triplet states approaches unity and the
first-order current becomes fully blocked. However, the
second-order processes are still allowed, leading to a finite
leakage current in the spin blockade regime. One can distinguish
different contributions coming from cotunneling:

(i) The double-barrier elastic second-order processes which
contribute directly to the current.

(ii) The double-barrier spin-flip cotunneling which contributes to
the current and reduces the occupation of the triplet state, this
way opening the DQD for the sequential tunneling.

(iii) The single-barrier spin-flip cotunneling which does not
contribute directly to the current but flips the spin in the DQD
and allows the sequential processes to occur.

\begin{figure}[t]
  \includegraphics[width = 0.68\columnwidth]{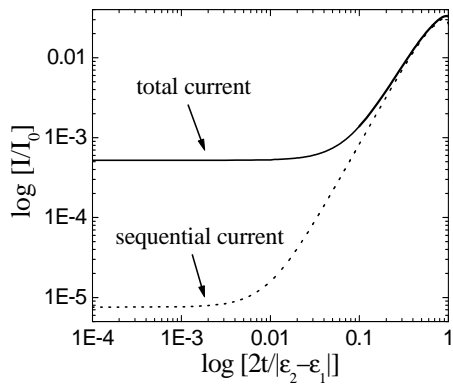}
  \caption{\label{Fig:8}
  The total (sequential plus cotunneling) current (solid line)
  and the sequential current (dotted line)
  in the parallel configuration
  as a function of the hopping between the two dots
  calculated in the middle of the Pauli spin blockade regime, $V=1$ mV.
  The parameters are the same as in Fig.~\ref{Fig:7}.
  The dependence of the current on $2t/|\varepsilon_2-\varepsilon_1|$
  in the antiparallel configuration is qualitatively similar.}
\end{figure}

As a consequence, in the Pauli spin blockade regime the current
flows due to cotunneling and spin-flip cotunneling-assisted
sequential tunneling. The key role is played by the spin-flip
processes which decrease the occupation of the two-electron
triplet states and lead to a finite occupation of singlet states.
The importance of the spin-flip processes in explanation of the
finite leakage current has been invoked in
Ref.~[\onlinecite{inarreaPRB07}]. The leakage was there associated
with spin-flip processes induced by hyperfine interaction through
the Overhauser effect. Here, we show that the leakage results just
from the interplay between different tunneling processes, i.e.
just from the pure nature of tunneling processes. We also note
that the effects of cotunneling are rather independent of the
material from which the dots are built, unlike the Overhauser
field. \cite{hansonRMP07} Thus, for example, in GaAs DQDs the
leakage current in the Pauli spin blockade would be associated
with both cotunneling and hyperfine contributions, while in carbon
nanotube DQDs it would be mainly due to cotunneling. Furthermore,
as shown in experiments by Ono and Tarucha, \cite{onoPRL04} who
studied the dependence of the leakage current on the applied
magnetic field, the hyperfine interaction starts to play a role at
certain finite in-plane magnetic fields, at which the nuclei
become polarized, leading to a sudden jump of the leakage current
as a function of magnetic field. This suggests that the main
contribution to the current in the spin blockade regime in the
absence of magnetic field may come from cotunneling. In the weak
coupling regime, typical values of the dot-lead coupling strength
$\Gamma$ are of the order of $\mu$eV, \cite{koganPRL04} which
gives the leakage current of the order of $10^{-3}I_0 \approx 1$
pA (see Fig.~\ref{Fig:8}), in agreement with experimental results.
\cite{onoPRL04} Finally, we notice that to explain the jump in the
leakage current when sweeping magnetic field, besides cotunneling,
one also has to include the hyperfine interaction. Nevertheless,
this goes beyond the scope of the present paper.

In Fig.~\ref{Fig:8} we present the logarithmic dependence of the
sequential and total currents in the parallel configuration on the
hopping parameter $t$. It is clearly visible that for
$2t/|\varepsilon_2 -\varepsilon_1| \ll 1$, the sequential current
is smaller by two orders of magnitude than the total current. On
the other hand, when $2t/|\varepsilon_2-\varepsilon_1| \approx 1$,
the sequential current becomes of the same order as the total
current, but still, noticeably, the total current is larger than
that calculated using only the first-order tunneling processes.
The saturation of the two currents for $2t/|\varepsilon_2
-\varepsilon_1| \ll 1$, see Fig.~\ref{Fig:8}, is associated with a
finite temperature. However, because sequential tunneling depends
exponentially on temperature while cotunneling only algebraically,
at lower $T$ the difference between the two currents would be even
more pronounced. The dependence of the current in the antiparallel
configuration is qualitatively similar to the one shown in
Fig.~\ref{Fig:8}. We also note that generally the above mentioned
mechanism responsible for the leakage current does not depend on
whether the leads are made of ferromagnetic or nonmagnetic
material.

The TMR as a function of the bias voltage is displayed in
Fig.~\ref{Fig:7}(b). One can see that the TMR calculated using the
first-order processes and that calculated taking into account
cotunneling are generally similar, except for the Pauli spin
blockade regime. In this transport regime the sequential TMR is
much overestimated. To understand this behavior we note again that
the blockade results from the full occupation of DQD's triplet
states. The probability is equally distributed between the three
components of the triplet in the parallel configuration. However,
in the antiparallel configuration, it turns out that the charge is
mainly accumulated in the $S_z = 1$ component of the triplet. As a
consequence, in the parallel configuration all the three
components of the triplet participate in transport, while in the
antiparallel configuration only one. Because the sequential
current in the blockade regime is mainly associated with thermal
fluctuations, the difference in the number of states relevant for
transport leads to large sequential TMR in blockade regime. This
difference also gives rise to the corresponding difference between
Fano factors calculated in the sequential tunneling approximation,
see the dotted curves in Fig.~\ref{Fig:7}(c). The shot noise is
then super-Poissonian and larger in the parallel configuration.
However, as pointed above, the leakage current in the Pauli spin
blockade regime is due to cotunneling and cotunneling-assisted
sequential tunneling processes. The total current flows not only
due to thermal fluctuations but due to correlated tunneling
through virtual states of the DQD. This fact generally decreases
the current fluctuations and the difference between the two
magnetic configurations. As a result, the total TMR and Fano
factor become suppressed as compared to the sequential tunneling
results, see Fig.~\ref{Fig:7}(b) and (c), respectively. In
addition, we also notice that, irrespective of the magnetic
configuration of the system, the shot noise is super-Poissonian in
the spin blockade regime and drops to sub-Poissonian value out of
the blockade regime, see Fig.~\ref{Fig:7}(c). The super-Poissonian
shot noise is generally due to bunching of spin-flip cotunneling
and cotunneling-assisted sequential tunneling processes.

\section{Double quantum dots coupled in parallel}

In this section we analyze the spin-polarized transport through
double quantum dots coupled in parallel. This geometry can be
realized by setting $\Gamma_{{\rm r}j} \equiv \Gamma/2$, for
$r={\rm L,R}$ and $j=1,2$, see Fig.~\ref{Fig:1}. We note that in
the case of negligible hopping between the two quantum dots, $t\to
0$, the behavior of the DQD system resembles that of a single
multi-level quantum dot. The problem of spin-dependent transport
through multi-level quantum dots has been addressed very recently
in Ref.~[\onlinecite{weymannPRB08}] and will not be considered
here.

\begin{figure}[t]
  \includegraphics[width = 0.7\columnwidth]{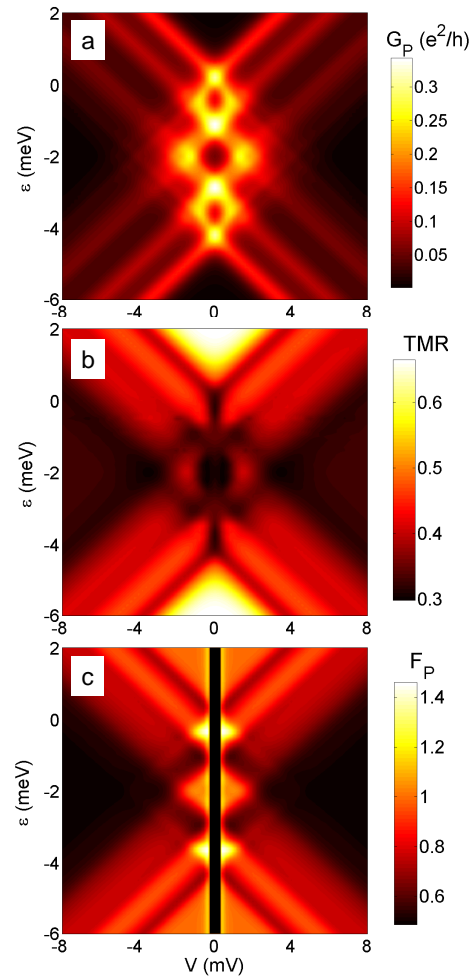}
  \caption{\label{Fig:9} (Color online)
  The differential conductance (a)
  and Fano factor (c) in the parallel magnetic configuration,
  and the TMR (b) as a function of the bias voltage $V$
  and the position of the dots' levels
  $\varepsilon \equiv \varepsilon_1 = \varepsilon_2$
  for double quantum dots coupled in parallel.
  The parameters are the same as in Fig.~\ref{Fig:2} with
  $\Gamma_{{\rm r}j} \equiv \Gamma/2$, for $r={\rm L,R}$, $j=1,2$,
  and $\Gamma=0.1$ meV. The transport regime where
  the Fano factor is divergent
  is marked with a thick black line.}
\end{figure}

In Fig.~\ref{Fig:9} we present the density plots of the
differential conductance and Fano factor in the parallel
configuration as well as of the TMR effect as a function of the
bias and gate voltages. The behavior of the conductance and the
shot noise in the antiparallel configuration is qualitatively
similar as in the parallel configuration, therefore, it is not
shown here. The information about the difference in transport in
the two magnetic alignments is contained in the TMR, whose
magnitude reflects the asymmetry in tunneling processes when the
leads are parallel or antiparallel to each other.

The differential conductance, shown in Fig.~\ref{Fig:9}(a),
displays characteristic Coulomb diamonds. Because the internal
energy structure of the DQD is generally the same as in the case
of DQD coupled in series, the differential conductances are
qualitatively similar, see Figs.~\ref{Fig:2}(a) and
\ref{Fig:9}(a). The main difference is that in the case of
parallel DQDs the level of each dot is coupled both to the left
and right leads, which generally results in an enhanced
conductance for parallel DQDs as compared to DQDs coupled in
series. Similarly the TMR shown in Fig.~\ref{Fig:9}(b), its bias
and gate voltage dependence is only slightly modified as compared
to Fig.~\ref{Fig:3}(a). Again, the linear TMR exhibits a strong
dependence on the occupation number of the DQD, with TMR given by
the Julliere formula for empty and fully occupied double quantum
dot, and is much suppressed in other transport regimes due to
spin-flip cotunneling. In the Coulomb blockade regime, the
transport properties are mainly conditioned by the spin-flip
cotunneling, which gives a dominant contribution to the current.
Furthermore, the bunching of spin-flip cotunneling in the Coulomb
blockade regime leads to super-Poissonian shot noise, see
Fig.~\ref{Fig:9}(c). The super-Poissonian shot noise is however
more pronounced in Coulomb blockade regimes with an odd number of
electrons in the double dot. On the other hand, for voltages above
the threshold for sequential tunneling, the shot noise becomes
sub-Poissonian, with the Fano factor approaching $1/2$.

Summing up, we notice that generally in the weak coupling regime
the effect of different geometries of the double dot system leads
only to qualitative difference in transport properties. The main
difference is in the magnitude of conductance -- for parallel DQDs
it is roughly two times larger than for serial DQDs. This is
contrary to the strong coupling regime where, for example, in the
parallel geometry the orbital Kondo phenomenon arises due to the
interference effects, while for DQDs coupled in series the orbital
Kondo effect is destroyed. \cite{blickPRB04} Finally, we also note
that in the case of detuned levels, $|\varepsilon_2
-\varepsilon_1| \gg 0$, transport properties of double quantum
dots coupled in parallel are qualitatively similar to those of
single multi-level quantum dots. \cite{weymannPRB08}

\section{T-shaped double quantum dots}

An interesting situation occurs when only one of the two quantum
dots is coupled to external leads, while the other one is
decoupled. In such T-shaped systems transport takes place through
the molecular states of the double dot, although only the first
dot is directly coupled to the leads. This may lead to new
transport behavior, especially to a large super-Poissonian shot
noise and an enhanced TMR, as we show in the sequel.

\begin{figure}[t]
  \includegraphics[height=13cm]{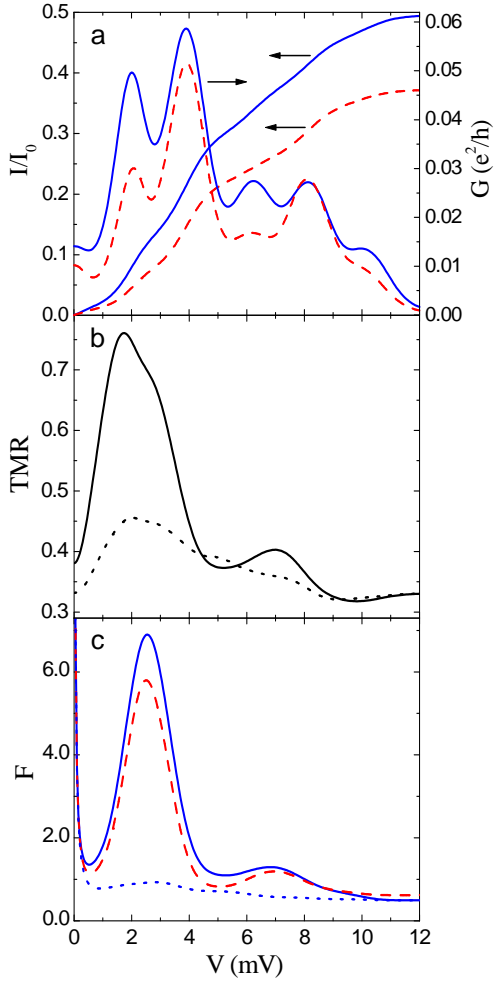}
  \caption{\label{Fig:10} (Color online)
  The bias dependence of the current and differential conductance
  (a), the TMR (b), and the Fano factor (c) in the
  parallel (solid line) and antiparallel (dashed line) configurations
  for T-shaped double quantum dots.
  The parameters are:
  $\Gamma_{{\rm r}1} \equiv \Gamma/2$, and
  $\Gamma_{{\rm r}2} = 0 $ for $r={\rm L,R}$,
  with $\Gamma=0.1$ meV, and $\e_1 = 1$ meV, $\e_2 = 0$ meV,
  $k_{\rm B} T=0.18$ meV, $U=2$ meV, $U^\prime=1$ meV,
  $t=0.2$ meV, and $p=0.5$.
  The dotted curves present the TMR (b) and
  the Fano factor in the parallel configuration (c)
  calculated for $\Gamma_{{\rm r}2} = \Gamma_{{\rm r}1}$.
  }
\end{figure}

\begin{figure}[t]
  \includegraphics[height=13cm]{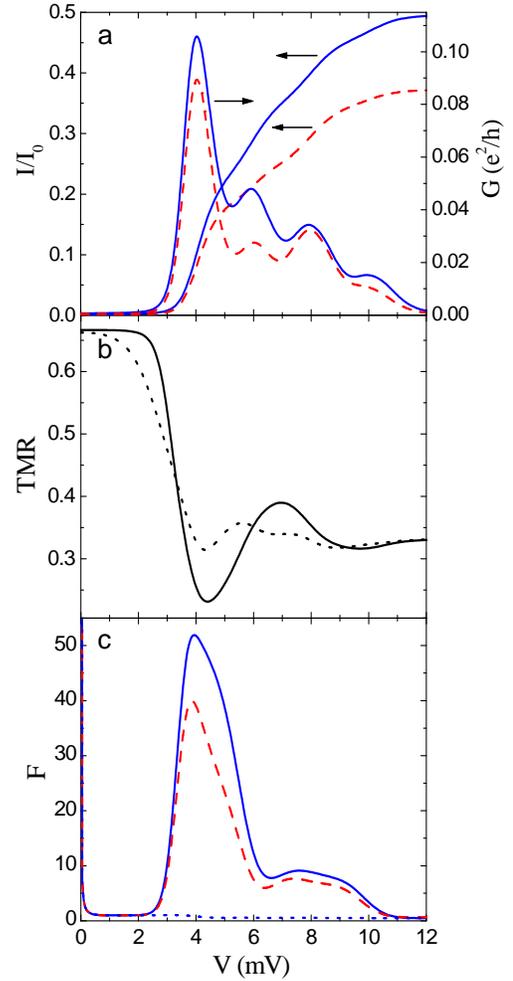}
  \caption{\label{Fig:11} (Color online)
  The bias dependence of the current, differential conductance
  (a), TMR (b) and the Fano factor (c) in the
  parallel (solid line) and antiparallel (dashed line) configurations.
  The parameters are the same as in
  Fig.~\ref{Fig:10} with $\e_2 = -4$ meV.
  The dotted curves present the TMR (b) and
  the Fano factor in the parallel configuration (c)
  calculated for $\Gamma_{{\rm r}2} = \Gamma_{{\rm r}1}$.
  }
\end{figure}

In Fig.~\ref{Fig:10} we present the current, differential
conductance, the TMR and the Fano factor as a function of the bias
voltage for the case when the first dot is coupled to the leads
while the second dot is decoupled. Spin-dependent transport
through single quantum dots have already been extensively studied.
However, the hopping $t$ between the two dots may modify the
transport properties significantly, giving rise to novel behavior.
First of all, we note that due to the finite $t$, transport takes
place through molecular states of the double dot. This in turn
leads to additional steps in the i-v curves and peaks in the
differential conductance, see Fig.~\ref{Fig:10}(a), as compared to
tunneling through one single-level quantum dots, where for
spin-degenerate level the current exhibits only two steps. The
difference between the currents in the parallel and antiparallel
configurations gives rise to the TMR which is shown in
Fig.~\ref{Fig:10}(b). The TMR displays an oscillatory-like
behavior with increasing the bias voltage. In addition, at the
threshold for sequential tunneling, the TMR becomes increased
above the Julliere TMR. \cite{julliere75} This result is rather
counterintuitive, as the Julliere TMR is characteristic of a
single ferromagnetic tunnel junction and one would expect that an
additional object (quantum dot) between the leads should decrease
the TMR. This was shown for single quantum dots with spin
degenerate levels and symmetric couplings to the leads where the
TMR was found to take at most the Julliere value.
\cite{weymannPRB05} In principle, the TMR can be enhanced above
the Julliere value when only one spin component of the dot takes
part in transport. Such spin selection may be achieved upon
applying an external magnetic field or due to finite exchange
interaction between spins in the dot.
\cite{weymannQDs,weymannJPCM07} In the case of T-shaped DQDs, we
find that the enhancement of the TMR is associated with increased
occupation of the second dot in the antiparallel configuration, as
compared to the parallel. Because the second dot is decoupled from
the leads, this increases the difference between the current in
the two configurations, yielding the TMR larger than the Julliere
value. In addition, finite occupation of the decoupled dot also
leads to large current fluctuations, as the rate for tunneling
between the second dot is much slower than that for tunneling
between the first dot and the leads. This in turn gives rise to
super-Poissonian shot noise, as displayed in Fig.~\ref{Fig:10}(c).
The super-Poissonian shot noise is present in both magnetic
configurations. It is also slightly larger in the parallel
configuration than in the antiparallel one, which is due to an
additional contribution to the noise coming from spin-dependent
tunneling between the dots and the leads. \cite{weymannPRB08} We
note that the super-Poissonian shot noise in T-shaped dots has
also been reported in the case of nonmagnetic leads.
\cite{djuricAPL05} The two dots were however rather weakly coupled
to each other so that no molecular states were formed, contrary to
the case considered here.

The enhancement of the TMR above the Julliere value and the large
super-Poissonian shot noise are present approximately at the
threshold for sequential tunneling. In this transport regime the
current is mainly mediated through the charge states of the
decoupled dot. The aforementioned effects should therefore become
washed out if there were a finite coupling between the second dot
and the leads. This is shown in Fig.~\ref{Fig:10}(b) and (c) where
we depict the TMR and Fano factor in the parallel configuration
calculated for $\Gamma_{{\rm r}2} = \Gamma_{{\rm r}1}$ (see the
dotted curves in Fig.~\ref{Fig:10}). The TMR larger than the
Julliere TMR accompanied by super-Poissonian shot noise is thus an
indication that transport takes place through quantum dot coupled
to another dot, which is not connected to the leads directly.

Figure~\ref{Fig:11} displays the bias dependence of the current,
TMR and the Fano factor calculated in the case when at equilibrium
the second dot is fully occupied while the first dot is empty. At
low bias voltage, the system is in the Coulomb blockade and the
double dot is occupied with two electrons in the second dot with
unit probability. The current flows then only due to elastic
cotunneling processes. This gives rise to the Julliere TMR and
Poissonian shot noise, see Fig.~\ref{Fig:11}(b) and (c). With
increasing the bias voltage, close to the threshold for sequential
tunneling, the occupation of the second dot is slightly lowered at
the cost of a finite occupation of another states. This opens the
system for the cotunneling-assisted sequential and sequential
tunneling processes. Nevertheless, because the double dot is still
mainly occupied by two electrons on the decoupled dot, this gives
rise to extremely large current fluctuations and Fano factors, see
Fig.~\ref{Fig:11}(c). This super-Poissonian shot noise can be
considerably reduced once the second dot becomes coupled to the
leads, see the dotted curve in Fig.~\ref{Fig:11}(c).

\begin{figure}[t]
  \includegraphics[width = 0.7\columnwidth]{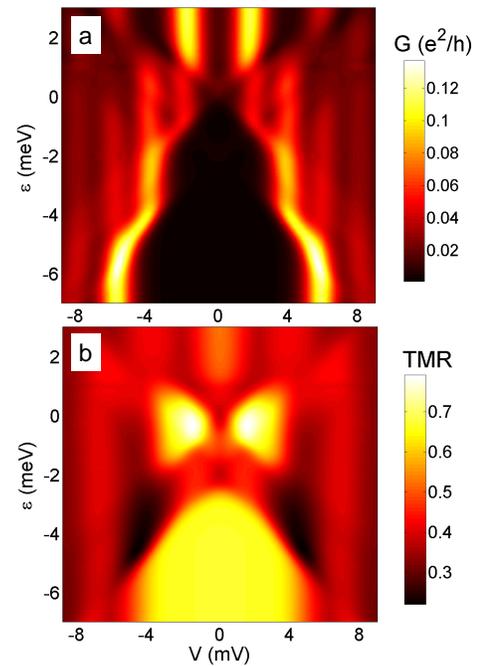}
  \caption{\label{Fig:12} (Color online)
  The differential conductance (a)
  in the parallel magnetic configuration,
  and the TMR (b) as a function of the bias voltage $V$
  and the position of the second dot level
  $\varepsilon \equiv \varepsilon_2$
  for T-shaped double quantum dots.
  The parameters are the same as in Fig.~\ref{Fig:10}.}
\end{figure}

Finally, in Fig.~\ref{Fig:12} we present the differential
conductance in the parallel configuration and the TMR as a
function of the bias voltage and the position of the second dot
level. The Coulomb blockade regimes are clearly visible in
Fig.~\ref{Fig:12}(a), whereas the different behavior of the TMR
depending on the transport regime is presented in
Fig.~\ref{Fig:12}(b).

\section{Conclusions}

We have analyzed the spin-dependent transport through systems
built of two strongly coupled quantum dots which are weakly
connected to external ferromagnetic leads. The considerations were
based on the real-time diagrammatic technique which allowed us to
determine transport properties in the sequential and cotunneling
regimes in a fully systematic way. In particular, we have analyzed
the current, differential conductance, shot noise and the TMR for
different geometries of the double dot system, including the
serial and parallel couplings, as well as the T-shaped systems. We
have also discussed the main differences in transport
characteristics corresponding to different DQDs geometries, which
may be helpful in determining the geometry of coupled quantum dots
in experiments.

In the case of double quantum dots coupled in series, we have
found an interesting dependence of the TMR on the occupation
number of the DQD. Furthermore, the super-Poissonian shot noise in
the Coulomb blockade regimes have been observed. On the other
hand, when the levels of the DQD were detuned, transport
characteristics revealed the Pauli spin blockade effects. We have
shown that the leakage current, observed experimentally in the
blockade regime, \cite{ono02} results from the interplay of
cotunneling processes which flip the spin in the DQD and make the
sequential tunneling possible. Thus, the current in the spin
blockade flows due to cotunneling and spin-flip
cotunneling-assisted sequential tunneling processes. This
mechanism is associated with pure nature of tunneling processes
and is thus relevant for both nonmagnetic and ferromagnetic leads.
In addition, we have also shown that the shot noise in the Pauli
spin blockade regime becomes super-Poissonian, while outside the
blockade it is sub-Poissonian.

For double quantum dots coupled in parallel, transport
characteristics were found to be qualitatively similar to those in
the case of DQDs coupled in series. The main difference is in the
magnitude of conductance -- for parallel DQDs it is roughly two
times larger than for serial DQDs.

In addition, we have also analyzed the case when the first quantum
dot is coupled to the leads while the second one is completely
decoupled. In such T-shaped systems we have found a large
super-Poissonian shot noise at the threshold for sequential
tunneling and the TMR enhanced above the Julliere value. These
effects are associated with an increased occupation of the
decoupled quantum dot, and become washed out once there is a
finite coupling between the second dot and the leads. Thus, the
enhanced TMR together with large super-Poissonian shot noise may
be an indication that transport takes place through quantum dot
which is side-coupled to another dot disconnected from the leads.


\begin{acknowledgments}

We acknowledge fruitful discussions with J. Barna\'s. This work
was supported by the Foundation for Polish Science and, as part of
the European Science Foundation EUROCORES Programme SPINTRA, by
funds from the Ministry of Science and Higher Education as a
research project in years 2006-2009.

\end{acknowledgments}


\appendix

\section{Details of numerical calculations}

In order to calculate the transport properties in the sequential
and cotunneling regimes, it is necessary to find the elements of
the respective first-order and second-order self-energy matrices.
This can be done using the respective diagrammatic rules.
\cite{thielmann,diagrams,weymannPRB05} Here, as an example, we
present the contribution coming from
$W_{\chi(N),\chi^\prime(N)}^{(2)}$ where $N$ is the charge state
of the double dot. It is given by
\begin{widetext}
\begin{eqnarray}
  W_{\chi(N),\chi^\prime(N)}^{(2)} &=& -2\pi \sum_{r,r'}\sum_{j,j'}
  \sum_{\sigma,\sigma'}\sum_{\chi''}\left\{
    |\matel{\chi}{U^\dagger d_{j\sigma} U}{\chi''}|^2
    |\matel{\chi''}{U^\dagger d_{j'\sigma'}^\dagger U}{\chi'}|^2
    \left[
      \gamma_{rj}^{\sigma+}(\e_{\chi''}-\e_\chi)B_{2r'j'}^{\sigma'-}(\e_{\chi''}-\e_{\chi'})
    \right.\right.\nonumber\\
      &&\left.+\gamma_{r'j'}^{\sigma'-}(\e_{\chi''}-\e_{\chi'})B_{2rj}^{\sigma+}(\e_{\chi''}-\e_{\chi})
      -B_{2rjr'j'}^{\sigma+\sigma'-}(\e_{\chi''}-\e_{\chi},\e_{\chi'}-\e_{\chi})
    \right]\nonumber\\
    &&+|\matel{\chi}{U^\dagger d_{j\sigma}^\dagger U}{\chi''}|^2
    |\matel{\chi''}{U^\dagger d_{j'\sigma'} U}{\chi'}|^2
    \left[
      \gamma_{rj}^{\sigma-}(\e_{\chi}-\e_{\chi''})B_{2r'j'}^{\sigma'+}(\e_{\chi'}-\e_{\chi''})
    \right.\nonumber\\
      &&\left.+\gamma_{r'j'}^{\sigma'+}(\e_{\chi'}-\e_{\chi''})B_{2rj}^{\sigma-}(\e_{\chi}-\e_{\chi''})
      -B_{2rjr'j'}^{\sigma-\sigma'+}(\e_{\chi}-\e_{\chi''},\e_{\chi}-\e_{\chi'})
    \right]\nonumber\\
    &&+\sum_{\chi'''}
    \matel{\chi}{U^\dagger d_{j\sigma}^\dagger U}{\chi'''}
    \matel{\chi'''}{U^\dagger d_{j'\sigma'} U}{\chi'}
    \matel{\chi}{U^\dagger d_{j'\sigma'} U}{\chi''}
    \matel{\chi''}{U^\dagger d_{j\sigma}^\dagger U}{\chi'}
    \times\nonumber\\
    &&\frac{2}{\e_{\chi}+\e_{\chi'}-\e_{\chi''}-\e_{\chi'''}}\left.
    \left[
      B_{1rjr'j'}^{\sigma-\sigma'+}(\e_{\chi''}-\e_{\chi'},\e_{\chi}-\e_{\chi'})
      -B_{1rjr'j'}^{\sigma-\sigma'+}(\e_{\chi}-\e_{\chi'''},\e_{\chi}-\e_{\chi'})
    \right]
  \right\} \,,
\end{eqnarray}
\end{widetext}
where $\gamma_{rj}^{\sigma\pm}(\e) = \Gamma_{rj}^\sigma
f^\pm(\e-\mu_r)/(2\pi)$, with $f^+$ being the Fermi distribution
function, $f^- = 1 - f^+$, $\mu_r$ denoting the electrochemical
potential of the lead $r$, and
\begin{eqnarray}
  B_{2rj}^{\sigma\pm}(\e) &=&
  \int d\omega
  \frac{\gamma_{rj}^{\sigma\pm}(\omega)}{(\omega-\e)^2}\nonumber\,,\\
  B_{\eta rjr'j'}^{\sigma\pm\sigma'\mp}(\e,\e') &=&
  \int d\omega \gamma_{rj}^{\sigma\pm}(\omega)\gamma_{r'j'}^{\sigma'\mp}(\omega-\e')
  \frac{1}{(\omega-\e)^\eta}\nonumber\,.
\end{eqnarray}
We note that having found all the first-order and second-order
self-energy matrices, we are in principle able to calculate
transport through arbitrary number of different orbital levels
coupled to each other and to external leads in an arbitrary way.
As far as numerical details are concerned, for systems consisting
of larger number of orbital levels, in calculations we make use of
the block structure of the initial Hamiltonian in the charge
space, and perform unitary transformation of the Hamiltonian and
the relevant local operators for each block separately. In
addition, we also store the respective matrix elements in blocks
labelled by charge quantum numbers.


\end{document}